\documentclass[12pt]{iopart}
\newcommand{\Yb}{\ensuremath{^{171}\mathrm{Yb}^+~}}
\newcommand{\rhouu}{\rho_{\uparrow\uparrow}}
\newcommand{\rhodu}{\rho_{\downarrow\uparrow}}
\usepackage{braket}
\usepackage{graphicx}
\usepackage{units}
\begin{document}

\title{Operational effects of the UNOT gate on classical and quantum correlations}

\author{
  Kuan Zhang,$^{1\ast}$
  Jiajun Ma,$^{1,2}$
  Xiang Zhang,$^{3,1}$
  Jayne Thompson,$^4$
  Vlatko Vedral,$^{2,4,5,1}$
  Kihwan Kim,$^{1\ast}$
  Mile Gu$^{6,7,4,1\ast}$
}

\address{
  $^1$
  Center for Quantum Information,
  Institute for Interdisciplinary Information Sciences,
  Tsinghua University,
  Beijing, P. R. China
}
\address{
  $^2$
  Department of Atomic and Laser Physics,
  Clarendon Laboratory,
  University of Oxford,
  Oxford OX1 3PU, United Kingdom
}
\address{
  $^3$
  Department of Physics,
  Renmin University of China,
  Beijing, P. R. China
}
\address{
  $^4$
  Centre for Quantum Technologies,
  National University of Singapore,
  Singapore
}
\address{
  $^5$
  Department of Physics,
  National University of Singapore,
  Singapore
}
\address{
  $^6$
  School of Mathematical and Physical Sciences,
  Nanyang Technological University,
  Singapore
}
\address{
  $^7$
  Complexity Institute,
  Nanyang Technological University,
  Singapore
}
\ead{
  \mailto{zhangkuan13@gmail.com},
  \mailto{kimkihwan@mail.tsinghua.edu.cn},
  \mailto{cqtmileg@nus.edu.sg}
}
\vspace{10pt}

\begin{abstract}
  The NOT gate that flips a classical bit is ubiquitous in classical information processing. However its quantum analogue, the universal NOT (UNOT) gate that flips a quantum spin in any alignment into its antipodal counterpart is strictly forbidden. Here we explore the connection between this discrepancy and how UNOT gates affect classical and quantum correlations. We show that while a UNOT gate always preserves classical correlations between two spins, it can non-locally increase or decrease their shared discord in ways that allow violation of the data processing inequality. We experimentally illustrate this using a multi-level trapped \Yb ion that allows simulation of anti-unitary operations.
\end{abstract}

%
\vspace{2pc}
\maketitle
%
%

\section{Introduction}

When given a quantum spin pointing in some unknown direction $\vec{n}$, is it possible to engineer a universal device that flips this spin to point in the antipodal direction $-\vec{n}$? While this process is easy to envision for classical vectors, it is strictly impossible for quantum spins. The quantum operation that takes an arbitrary quantum state $\ket\varphi$ to its orthogonal complement $\ket{\varphi^\bot}$ is anti-unitary, and thus does not exist~\cite{Gisin99,Bechmann99,Werner99}. Like the no-cloning theorem, this uniquely quantum constraint has drawn significant scientific interest~\cite{Martini02,Ricci04,Scarani05}.

In contrast to cloning, the radical operational consequences of the UNOT gate are not as readily apparent on a single qubit. Suppose Alice secretly encodes a direction $\vec{n}$ in 3-dimensional space by preparing a spin aligned in $\vec{n}$. She then challenges Bob to estimate $\vec{n}$. If Bob can perfectly clone quantum states, then he can violate the uncertainty principle by measuring each clone in a different complementary basis. On the other hand, any measurement Bob makes after applying a UNOT gate on the input spin can be simulated by measuring the input directly and reinterpreting the measurement outcome (recording `up' as `down' and vice versa). Thus UNOT gates do not allow Bob to retrieve information about $\vec{n}$ beyond standard quantum limits.

The consequences of the UNOT gate surface when an ancillary qubit is introduced. Consider the same game, but now played on two qubits. Instead of sending a single spin, Alice now sends a pair of spins. Take two different strategies, either (a) sending Bob two aligned spins, both in direction $\vec{n}$, or (b) an anti-aligned pair, with one spin in direction $\vec{n}$, and the other in direction $-\vec{n}$. Gisin and Popescu illustrated that the second strategy improves Bob's capacity to estimate $\vec{n}$~\cite{Gisin99,Jeffrey06}. They noted that if Bob possesses a UNOT gate, he can deterministically convert a pair of aligned spins to anti-aligned spins, and thus break standard quantum limits whenever Alice adopts strategy (a).

\begin{figure}
  \centering
  \includegraphics[width=0.8\columnwidth]{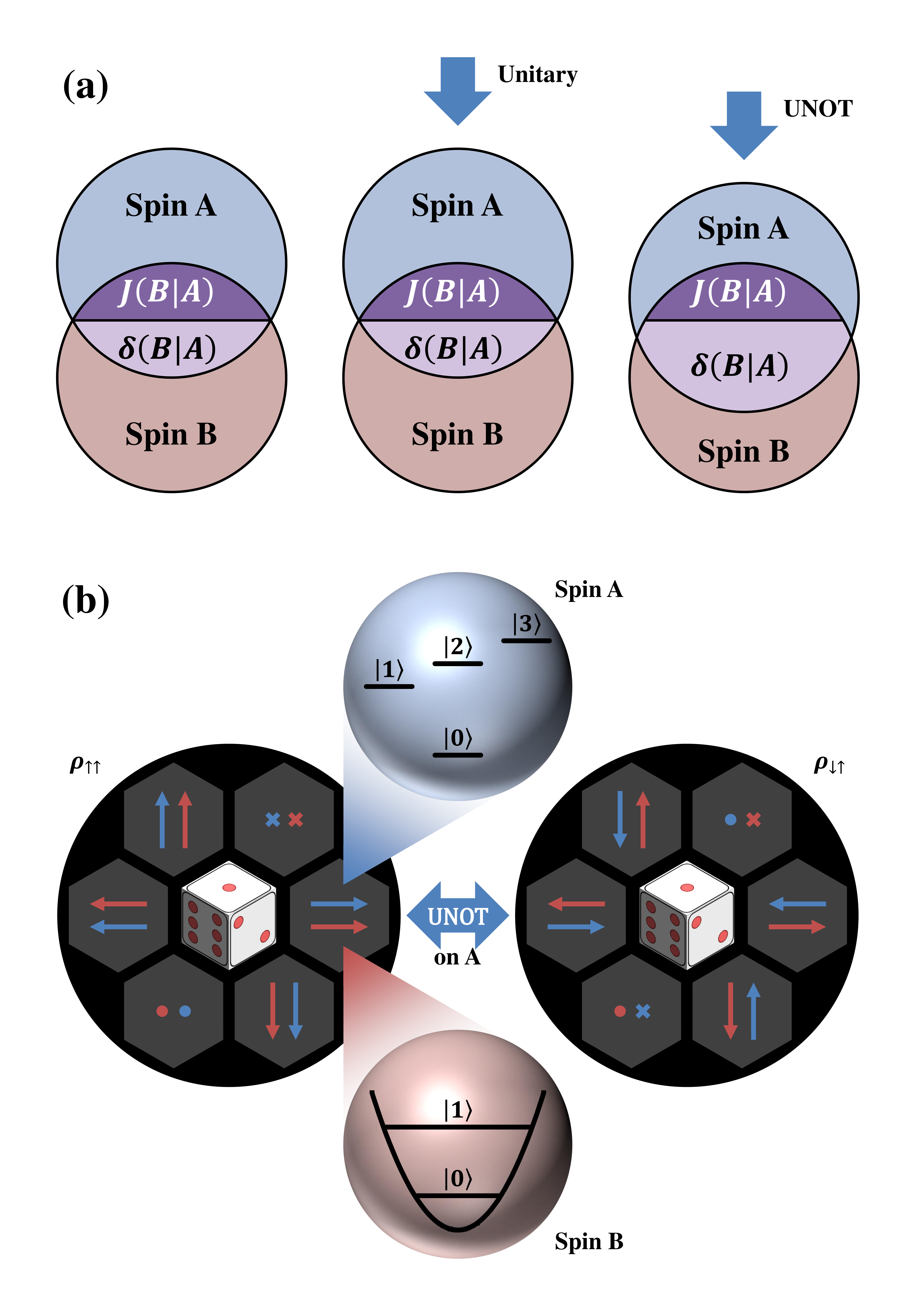}
  \caption{
    \textbf{UNOT gates on spin pairs.}
    (a) Local reversible operations leave classical correlations $J(B|A)$ and quantum correlations $\delta(B|A)$ unchanged, where $A$ and $B$ are two separable spins. A UNOT gate also preserves $J(B|A)$, but can change $\delta(B|A)$.
    (b) We simulate the effect of UNOT on $\rhouu$ and $\rhodu$, where the two spins are encoded in the internal and external degrees of freedom of a trapped \Yb ion in a harmonic potential. Spin $A$ is mapped to a 4-level system using Eq.~(\ref{eq:M}), spanned by the basis
    $\ket0_{\rm{A}}=
     \ket{F=0,m_{\rm{F}}=0}
    $ and
    $\ket{n=1,2,3}=
     \ket{F=1,m_{\rm{F}}=n-2}
    $,
    where $F$ and $m_{\rm{F}}$ characterize the total internal angular momentum of \Yb. The transition frequency from $\ket{F=0,m_{\rm{F}}=0}$ to $\ket{F=1,m_{\rm{F}}}$ is
    $(2\pi)
     (12642.8+9.0m_{\rm{F}})
     \unit{MHz}
    $.
    Spin $B$ is mapped to the ground and first excited states of external motional mode, denoted by $\ket0_{\rm{B}}$ and $\ket1_{\rm{B}}$, which are separated by the trap frequency
    $(2\pi)2.44
     \unit{MHz}
    $.
  }
  \label{fig:Pair}
\end{figure}

This connection suggests that UNOT gate fields distinctive effects on quantum correlations. Here we formalize this intuition using recent methods that isolate the purely quantum component of correlations between two systems. These correlations, known as discord, are often motivated as correlations accessible only to quantum observers~\cite{Ollivier01,Vedral01,MileGu12}. We show that the UNOT gate preserves classical correlations between two spins, but can change their shared quantum correlations in ways forbidden by fundamental data processing principles (see figure~\ref{fig:Pair}(a)). We illustrate this through experiment -- by adapting recent ion trap technology that allows perfect simulation of anti-unitary operations~\cite{Leibfried03,Casanova11,XiangZhang15}. We then outline how these results rationalize the discrepancy in communication rate between aligned and anti-aligned spins, showing that it exactly relates to the UNOT gate's non-trivial effect on discord during the decoding process.

\section{Theory}

\subsection{Technical framework}

Consider first two classical spins $A$ and $B$. Let $S(\cdot)$ denote the information entropy function, such that $S(A)$ and $S(B)$ quantify the respective uncertainties of $A$ and $B$ when viewed independently and $S(AB)$ the uncertainty of the joint spin pair. The mutual information
$I(A,B)=
 S(A)+S(B)-S(AB)
$
then captures the total correlations between $A$ and $B$. This coincides with
$J(B|A)=
 S(B)-S(B|A)
$,
the reduction of uncertainty in $B$, when someone measures and communicates the state of $A$. The \emph{data processing inequality} implies that $I(A,B)$ can never increase under local operations~\cite{Cover06,Nielsen10}. This reflects the principle that we can never spontaneously obtain more information about a spatially separated system $B$ through local operations on $A$. Consequently, any reversible operation on $A$ must conserve $I(A,B)$ -- a condition clearly satisfied by the classical NOT gate.

When the two spins are quantum, the analogues of $I(A,B)$ and $J(B|A)$ no longer coincide. Any positive operator valued measurements (POVMs) $\{\Pi_a\}$ on $A$ can induce unavoidable noise, limiting the entropy reduction on $B$ to
$J(B|A)=
 \sup_{\{\Pi_a\}}
 \left[
   S(B)-
   \sum_a{
     p_aS(B|a)
   }
 \right]
$,
where $p_a$ is the probability of getting outcome $a$, and $S(B|a)$ represents the corresponding entropy of $B$ conditioned on this outcome. In literature, $J(B|A)$ is considered to be the classical component of $I(A,B)$, as it represents how much information a \emph{classical observer} can gain about $B$ when measuring $A$. The remaining portion,
$\delta(B|A)=
 I(A,B)-J(B|A)
$,
is defined as the quantum discord, and interpreted as the purely quantum correlations between $A$ and $B$~\cite{Ollivier01,MileGu12}.

\subsection{Theoretical results}

Let $\rho$ be a separable bipartite state on two spins, $A$ and $B$. Here we establish the following relations between the UNOT gate and classical and quantum correlations:
\begin{enumerate}
  \item[(i)] Local UNOT gates preserve $J(B|A)$ and $J(A|B)$.
  \item[(ii)] If $\delta(B|A)=0$, a local UNOT gate conserves $I(A,B)$.
  \item[(iii)] Local UNOT gates can nevertheless violate the data processing inequality, but only when $\delta(B|A)>0$.
\end{enumerate}

Result (i) implies that the UNOT gate can never break the data processing inequality for classical correlations. A classical observer that quantifies correlations by local measurement (i.e., $J(A|B)$ or $J(B|A)$) will conclude that the UNOT gate has no radical effects. Meanwhile result (ii) implies that the UNOT gate always obeys the data processing inequality, provided no discord is present. Finally result (iii) shows that the UNOT gate can violate the data processing inequality, but only when discord is present. Furthermore, this violation can only be witnessed when purely quantum correlations are taken into account. Proofs for (i) and (ii) are given in \ref{sec:Proofs}, while we demonstrate (iii) directly by explicit examples.

\section{Experiment}

\subsection{Protocol}

Our experiment considers a separable bipartite quantum state $\rhodu$ constructed as follows: First let
$\mathcal{I}=
 \left\{
   \vec{x},
  -\vec{x},
   \vec{y},
  -\vec{y},
   \vec{z},
  -\vec{z}
 \right\}
$
represent a set of six standard coordinate directions in 3-dimensional space. We select a direction
$\vec{n}\in
 \mathcal{I}
$
uniformly at random, and prepare one spin aligned in direction $-\vec{n}$ and the other in $\vec{n}$. Denote their states respectively by $\ket{-\vec{n}}_{\rm{A}}$ and $\ket{\vec{n}}_{\rm{B}}$. The choice of $\vec{n}$ is then discarded. The resulting mixed state
$\rhodu=
 \sum_{
   \vec{n}\in
   \mathcal{I}
 }{
   \ket{-\vec{n}}_{\rm{A}}
   \bra{-\vec{n}}_{\rm{A}}
   \otimes
   \ket{\vec{n}}_{\rm{B}}
   \bra{\vec{n}}_{\rm{B}}
 }/6
$
then describes two anti-aligned spins that are oriented along one of the six directions in $\mathcal{I}$ at random. Application of the UNOT gate on one of the two spins then results in the state
$\rhouu=
 \sum_{
   \vec{n}\in
   \mathcal{I}
 }{
   \ket{\vec{n}}_{\rm{A}}
   \bra{\vec{n}}_{\rm{A}}
   \otimes
   \ket{\vec{n}}_{\rm{B}}
   \bra{\vec{n}}_{\rm{B}}
 }/6
$
which represents two aligned spins oriented randomly in some direction
$\vec{n}\in
 \mathcal{I}
$
(see figure~\ref{fig:Pair}(b)). By result (i), $\rhouu$ and $\rhodu$ must have coinciding classical correlations $J(B|A)$.

Here we conduct two separate experiments. In the first, we prepare $\rhodu$ on a spin pair, and simulate the action of the UNOT gate on one of the two spins. We characterize, by tomography, the effect of this action on the classical and quantum correlations within the spin pair. This process is then repeated with $\rhouu$ in place of $\rhodu$. This allows us to experimentally demonstrate result (iii) by showing that the local application of a UNOT gate can fundamentally increase or decrease $\delta(B|A)$ while preserving $J(B|A)$, and thus violate the data processing inequality.

\begin{figure*}
  \centering
  \includegraphics[width=\columnwidth]{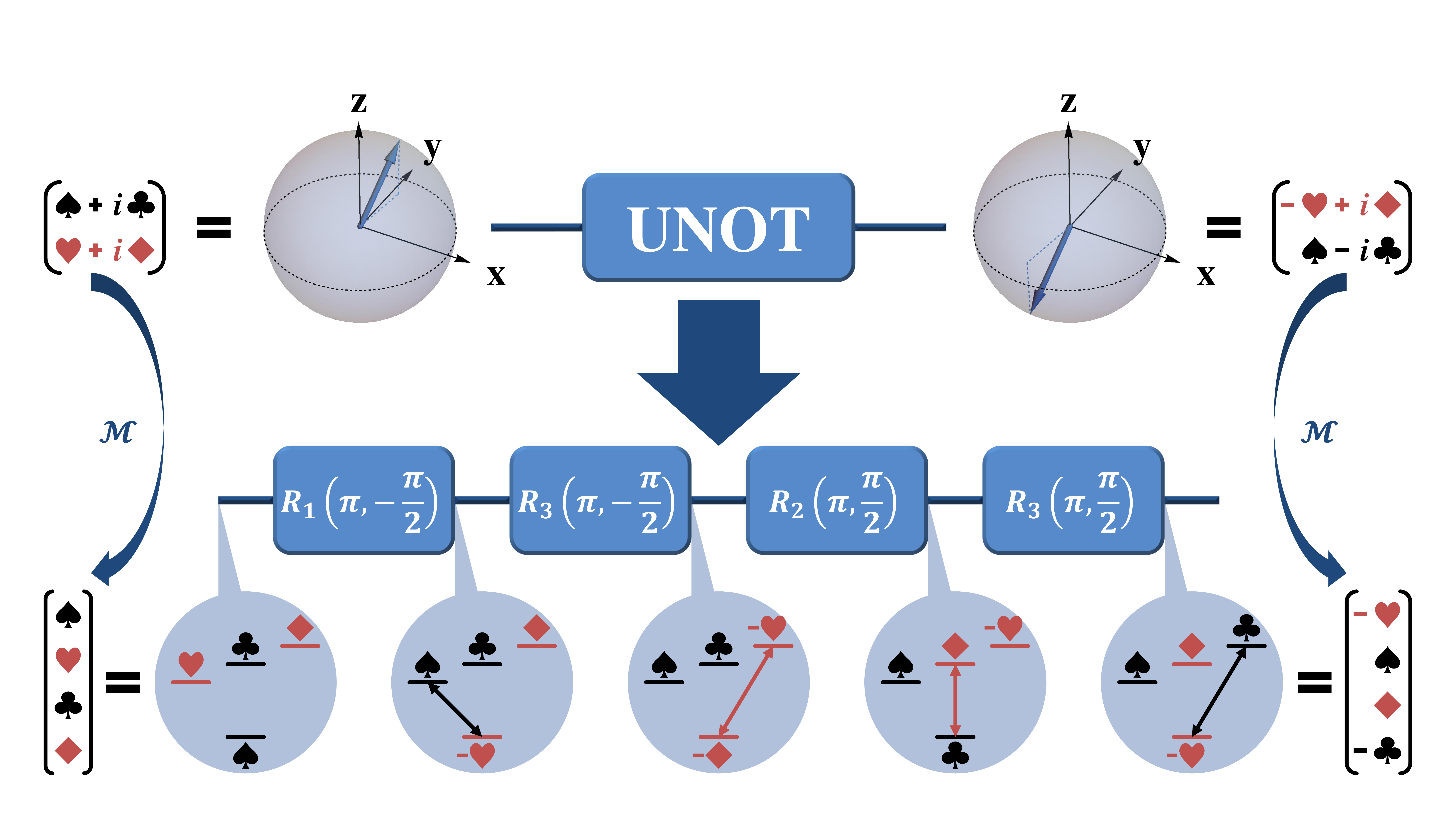}
  \caption{
    \textbf{Simulation of UNOT Gate on encoded spin A.}
    While the UNOT gate is unphysical, it can be exactly simulated using a quantum 4-level system. To simulate UNOT acting on state
    $\ket\varphi=
     \alpha\ket\uparrow+
     \beta\ket\downarrow
    $,
    we initialize a suitable state
    $\ket{\bar\varphi}=
     \mathcal{M}
     \ket\varphi
    $
    on the 4-level system (see Eq.~\ref{eq:M}). The expected output state
    $\ket{\varphi^\bot}=
     \Theta_{\rm{UNOT}}
     \ket\varphi=
    -\beta^*\ket\uparrow+
     \alpha^*\ket\downarrow
    $
    can then be simulated by applying
    $\bar\Theta_{\rm{UNOT}}=
     \ket1\bra0-
     \ket0\bra1-
     \ket3\bra2+
     \ket2\bra3
    $
    on $\ket{\bar\varphi}$. It is easy to check that
    $\bar\Theta_{\rm{UNOT}}
     \ket{\bar\varphi}=
     \mathcal{M}
     \ket{\varphi^\bot}
    $.
    In the ion trap system, $\bar\Theta_{\rm{UNOT}}$ can be realized with 4 microwave pulses, where
    $R_n(\pi,\phi)=
     -i\left(
       e^{-i\phi}
       \ket{n}_{\rm{A}}
       \bra0_{\rm{A}}+
       \rm{h.c.}
     \right)+
     \sum_{m\neq0,n}{
       \ket{m}_{\rm{A}}
       \bra{m}_{\rm{A}}
     }
    $,
    $n=1,2,3$.
  }
  \label{fig:UNOT}
\end{figure*}

\subsection{Simulating UNOT}

The experiment involves exact simulation of the UNOT gate -- an unphysical operation. Indeed, all existing demonstrations of the UNOT gate are based on theoretically optimal approximations~\cite{Martini02,Ricci04}. We circumvent these issues by embedding the state of spin $A$ within a larger Hilbert space (see figure~\ref{fig:UNOT}) -- a technique recently proposed for exactly simulating anti-unitary operations~\cite{Casanova11,XiangZhang15}. This approach maps each spin state
$\ket\varphi=
 \alpha\ket\uparrow+
 \beta\ket\downarrow
$
to a corresponding state
\begin{equation}
  \label{eq:M}
  \mathcal{M}
  \ket{\varphi}=
  \ket{\bar\varphi}=
  \alpha_{\rm{R}}\ket0+
  \beta_{\rm{R}}\ket1+
  \alpha_{\rm{I}}\ket2+
  \beta_{\rm{I}}\ket3
\end{equation}
on some 4-level quantum system, where
$\alpha=
 \alpha_{\rm{R}}+
 i\alpha_{\rm{I}}
$ and
$\beta=
 \beta_{\rm{R}}+
 i\beta_{\rm{I}}
$,
and $\ket{n=0,1,2,3}$ denotes some orthogonal basis. The action of the UNOT gate can then be simulated by a suitable unitary operator on the 4-level system.

\subsection{Experimental realization}

We use a trapped \Yb ion in a harmonic potential~\cite{Leibfried03}. Spin $B$ is encoded within the ground and first excited external motional states of the ion, denoted $\ket0_{\rm{B}}$ and $\ket1_{\rm{B}}$. Meanwhile spin $A$ is encoded within the four internal degrees of freedom of the trapped ion using the aforemention technique (see figure~\ref{fig:Pair}(b)). Arbitrary unitary operations on this 4-level system can be implemented via a sequence of appropriate microwave pulses~\cite{XiangZhang15,YangchaoShen17}. This allows us to simulate any unitary or anti-unitary operation on spin $A$. In particular, we develop an explicit pulse sequence for simulating the UNOT gate as shown in figure~\ref{fig:UNOT}.

To determine the classical and quantum correlations between spins, we develop an efficient tomography scheme that directly reconstructs the $4\times4$ density operator of the spin pair. This technique allows us to avoid reconstructing the full $8\times8$ density operator necessary to describe the state of the trapped ion, significantly reducing the number of necessary measurements to reach set levels of accuracy (see \ref{sec:Tomo}). In addition, our approach involves synthesizing specialized interactions between the ion's internal and external degrees of freedom, different from the standard red sideband operation~\cite{Gulde03,Monz09,JunhuaZhang16}. These interactions allow direct Bell-basis measurements for the encoded spin pair (see \ref{sec:Ops}).

\begin{figure*}
  \centering
  \includegraphics[width=\columnwidth]{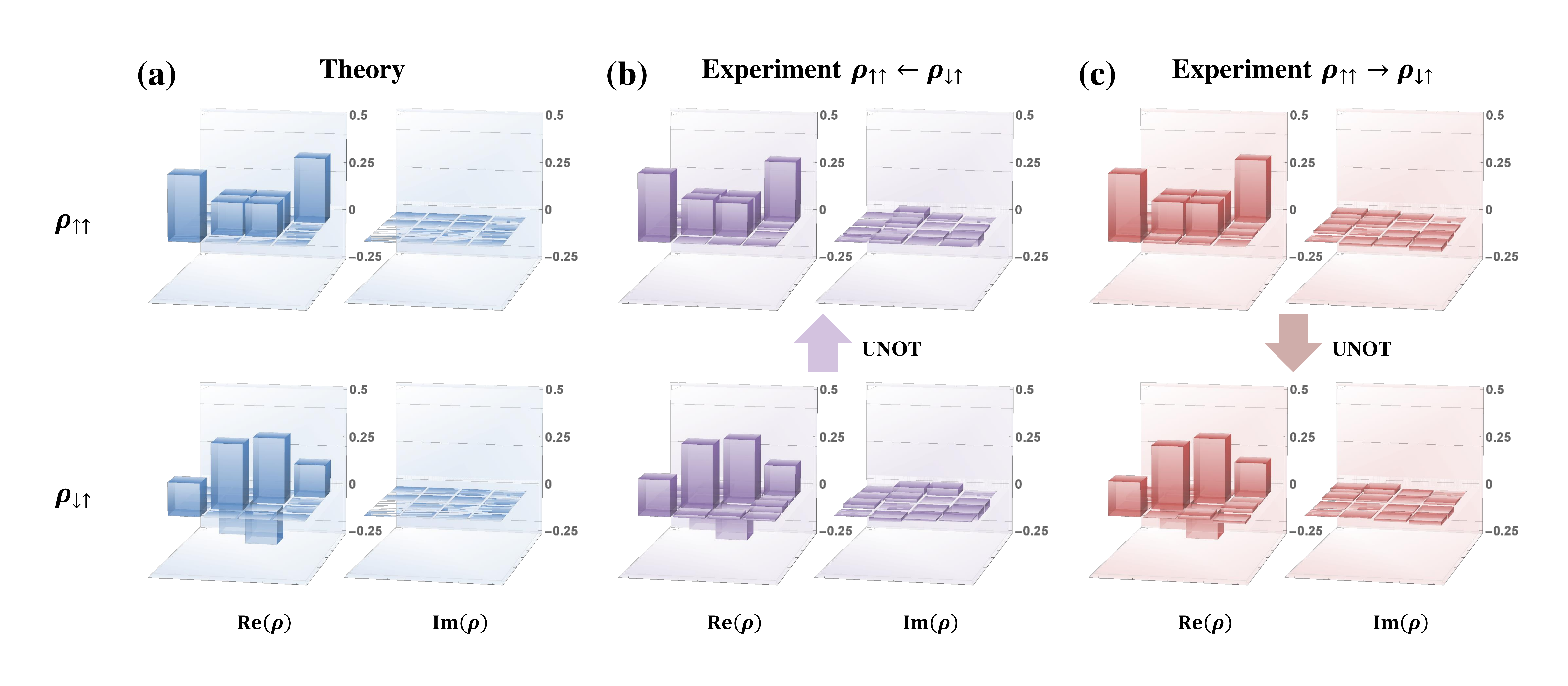}
  \caption{
    \textbf{Experimentally measured density operators compared with theoretical predictions.}
    (a) illustrates the theoretically predicted density operators for $\rhouu$ and $\rhodu$, where each vertical bar represents a corresponding matrix element.
    In the first experiment (b), $\rhodu$ is prepared with a fidelity of $0.992\pm0.004$. After simulating the UNOT gate, we retrieve $\rhouu$ with fidelity of $0.997\pm0.007$.
    In the second experiment (c), $\rhouu$ is prepared with fidelity $0.998\pm0.004$, and the final state approximates $\rhodu$ to fidelity $0.997\pm0.002$. The errors are estimated by Monte Carlo methods with a confidence level of 95\%.
  }
  \label{fig:DM}
\end{figure*}

\begin{figure}
  \centering
  \includegraphics[width=\columnwidth]{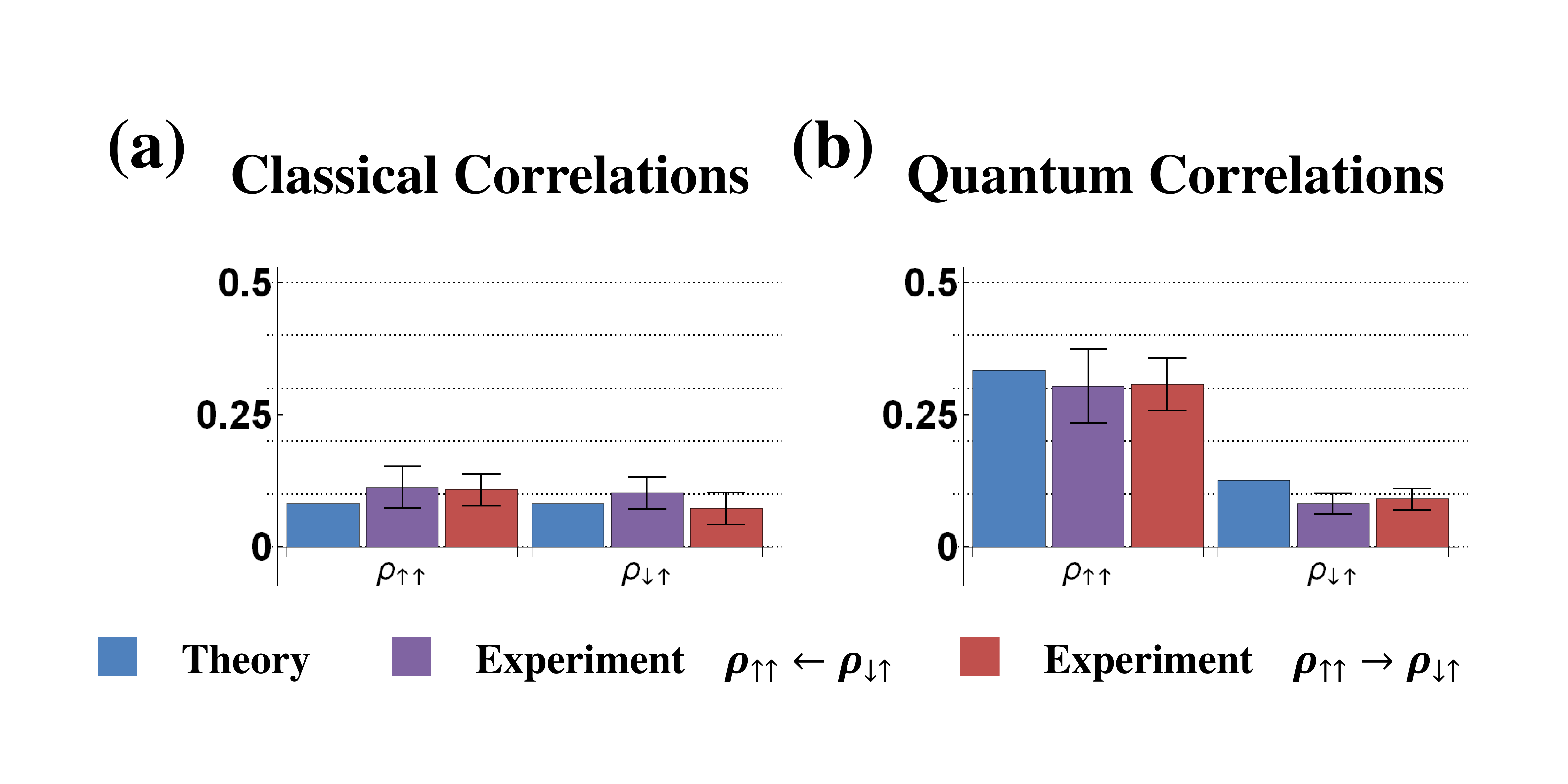}
  \caption{
    \textbf{The effect of UNOT gate on various types of correlations.}
    The theoretically predicted effect of UNOT gate (blue bars), together with experimentally measured effects when acting on $\rhodu$ (purple bars) and $\rhouu$ (red bars) are displayed for (a) classical correlations $J(B|A)$ and (b) quantum correlations $\delta(B|A)$.
    (a) Theory predicts that $J(B|A)=0.082$ for both $\rhouu$ and $\rhodu$. Experimental results agree within experimental error.
    (b) Theory predicts that $\delta(B|A)$ is $0.415$ for $\rhouu$ and $0.208$ for $\rhodu$ -- a difference of $0.207$. This agrees with experiment, where we see respective increase and decrease of $0.22\pm0.07$ and $0.22\pm0.05$ when converting to and from $\rhouu$.
  }
  \label{fig:Correlation}
\end{figure}

\subsection{Experimental results}

We conduct two separate experimental tests. The first initializes $\rhodu$, and determines how its classical and quantum correlations are affected by a UNOT operation on spin $A$. The second experiment repeats this process, using $\rhouu$ in place of $\rhodu$. In both cases, tomographic data is collected over 102000 trials (see \ref{sec:Prep} and \ref{sec:Tomo}). The experimentally reconstructed density operators agree with the theoretical predictions to high fidelity (see figure~\ref{fig:DM}).

The experimentally observed effects of the UNOT gate on classical and quantum correlations are illustrated in figure~\ref{fig:Correlation}. They confirm that the amount of classical correlations between $A$ and $B$ are preserved, in agreement with result (i). Meanwhile the application of UNOT on $\rhodu$ increases quantum correlations by $0.22\pm0.07$, violating the data processing inequality by over 5 standard deviations and thus establishing (iii). The application of the UNOT gate on $\rhouu$ can reverse this, inducing a decrease in $\delta(B|A)$ of $0.22\pm0.05$.

\section{Relation to communicating with aligned vs anti-aligned spins}

To conclude, we tie our results back to the communication advantage of using anti-aligned versus aligned spin pairs~\cite{Gisin99}. Specifically, consider the scenario where Alice encodes a classical message into two aligned spins, using the six possible codewords
$\left\{
   \ket{\vec{n}}_{\rm{A}}
   \ket{\vec{n}}_{\rm{B}}
 \right\}_{
   \vec{n}\in
   \mathcal{I}
 }
$
with equiprobability. The communication capacity is given by the Holevo quantity
$\chi_{\uparrow\uparrow}=
 S(\rhouu)
$,
which bounds the amount of information Alice can communicate per spin pair~\cite{Holevo73}. This bound can always be saturated in the independent and identically distributed (i.i.d) limit, but never exceeded by conventional means.

Bob, however, can use a UNOT gate to surpass this bound. The rationale being that the capacity of such a channel
$\chi_{\uparrow\uparrow}=
 S(A)+S(B)-I(A,B)
$
can be divided into $S(A)+S(B)$, representing the amount of information $A$ and $B$ each individually communicate discounting correlations; and $I(A,B)$, a correction term that captures how much of the aforementioned information is redundant.
$I(A,B)=
 J(B|A)+\delta(B|A)
$
can then be further divided into classical and quantum components. The UNOT gate, acting on $\rhouu$, can uniquely reduce $\delta(B|A)$, and thus reduce redundant information about $\vec{n}$ encoded in the two spins. The UNOT gate thereby boosts the amount of information Bob can extract. The performance gain is given by
$\Delta\chi=
 \delta_{\uparrow\uparrow}(B|A)-
 \delta_{\downarrow\uparrow}(B|A)
 \approx0.21
$,
where $\delta_{\uparrow\uparrow}$ and $\delta_{\downarrow\uparrow}$ respectively represent the discord of $\rhouu$ and $\rhodu$. This gain exactly coincides with the change in discord.

Observe also that Bob's application of the UNOT gate is functionally equivalent to Alice encoding the message in anti-aligned spins, i.e., using the codewords
$\left\{
   \ket{-\vec{n}}_{\rm{A}}
   \ket{\vec{n}}_{\rm{B}}
 \right\}_{
   \vec{n}\in
   \mathcal{I}
 }
$
in place of
$\left\{
   \ket{\vec{n}}_{\rm{A}}
   \ket{\vec{n}}_{\rm{B}}
 \right\}_{
   \vec{n}\in
   \mathcal{I}
 }
$.
Thus, we see that the performance advantage of having UNOT gates exactly coincides with the performance discrepancy between using aligned vs. anti-aligned spins. Note also that this advantage only exists assuming Bob can measure in an entangling basis. This corroborates recent evidence that many operational effects of discord can only be accessed via entangling measurements~\cite{MileGu12,Modi14,Weedbrook13}. That is, the UNOT only imparts unphysical effects on the quantum component of the correlations between two spins.

\section{Discussion and conclusion}

This article explores the UNOT gate's capacity to locally increase the correlations between spins, and thus break the data processing inequality. We establish that this gate shows no such capability in classical domain. It obeys the data processing inequality for all classically correlated systems. Furthermore, it preserves all classical correlations within general correlated quantum systems. Violation can only be witnessed when quantum correlations are explicitly considered. We adopt state of the art techniques for simulating anti-unitary operations to experimentally demonstrate this phenomena using a trapped \Yb ion. A violation of the data processing inequality by over 5 standard deviations is observed. These results connect the unphysicality of the UNOT gate, its effect on quantum correlations, and the discrepancy in communicating using aligned versus anti-aligned spins. Our experiment then highlights how such unphysical effects can be simulated using present day ion trap technology.

There are a number of directions in which these results can generalize. Observe that the UNOT gate is equivalent, up to local unitary rotation, to any other anti-unitary operator on a qubit. The effect of UNOT gates on classical and quantum correlations thus also applies to any other anti-unitary operation. Meanwhile we may extend these ideas to systems of higher dimensions. For example, in optical systems, one can communicate more information using a conjugate pair of coherent states than using the same state twice~\cite{Niset07}. This performance difference is likely due to an analogous effect of anti-unitary operations on bipartite Gaussian correlations.

Our results could also offer a new way to capture what classes of correlations are quantum. One could take as an axiom that states whose correlations change under the UNOT gate are quantum. The motivation being such effects field no classical explanation. Such an axiom would identify certain discordant states such as
$\rho=
 \left(
   \ket{
     \vec{x},
     \vec{x}
   }
   \bra{
     \vec{x},
     \vec{x}
   }+
   \ket{
     \vec{z},
     \vec{z}
   }
   \bra{
     \vec{z},
     \vec{z}
   }
 \right)/2
$
as more classical than others, since we can replicate the effect of a UNOT gate acting on them using only local unitary operations. This could well generalize to multi-partite systems, and present a general operational criterion for identifying quantum correlations residing somewhere between discord and entanglement.

\ack{
  This project was supported by the National Key Research and Development Program of China under Grants No. 2016YFA0301900 (No. 2016YFA0301901), and the National Natural Science Foundation of China Grants No. 11374178, and No. 11574002, the John Templeton Foundation Grant 53914 "Occam's Quantum Mechanical Razor: Can Quantum theory admit the Simplest Understanding of Reality?", the National Research Foundation of Singapore (NRF Award No. NRF--NRFF2016--02), the Competitive Research Programme (CRP Award No. NRF-CRP14-2014-02), the Ministry of Education in Singapore, and the Oxford Martin School.
}

\clearpage

\appendix

\section{
  Theoretical proofs
  \label{sec:Proofs}
}

\subsection{Proof of result (i)}

Note first that as $\rho$ is separable, it can be written in the form
$$\rho=
  \sum_i{
    p_i
    \ket{\varphi_i}_{\rm{A}}
    \bra{\varphi_i}_{\rm{A}}
    \otimes
    \ket{\phi_i}_{\rm{B}}
    \bra{\phi_i}_{\rm{B}}
  }.
$$
Note also that $J(B|A)$ is asymmetric. Thus we consider the cases where UNOT gate on spin $A$ and spin $B$ separately.

When the UNOT gate is applied to $A$, the state is transformed to
$$\rho^{\rm{o}}=
  \sum_i{
    p_i
    \ket{\varphi_i^\bot}_{\rm{A}}
    \bra{\varphi_i^\bot}_{\rm{A}}
    \otimes
    \ket{\phi_i}_{\rm{B}}
    \bra{\phi_i}_{\rm{B}}
  }
$$
where $\ket{\varphi_i^\bot}_{\rm{A}}$ satisfies
$\langle
   \varphi_i|
   \varphi_i^\bot
 \rangle_{\rm{A}}=0
$.
Let $J^{\rm{o}}(B|A)$ and $\delta^{\rm{o}}(B|A)$ denote the resulting classical and quantum correlations in $\rho^{\rm{o}}$. Recall also that the definition of $J(B|A)$ involves an optimization over the measurement basis for $A$. For a given rank-1 POVM, $\{\Pi_a\}$, consider the basis-dependent classical correlations
$J_{\{\Pi_a\}}(B|A)=
 S(B)-
 \sum_a{
   p_aS(B|a)
 }
$.
We now introduce a second rank-1 POVM, $\{\Pi_a^\bot\}$, whose projective operators satisfy $\tr_{\rm{A}}(\Pi_a\Pi_a^\bot)=0$. Thus
$\tr_{\rm{A}}
 \left(
   \Pi_a
   \ket\varphi_{\rm{A}}
   \bra\varphi_{\rm{A}}
 \right)=
 \tr_{\rm{A}}
 \left(
   \Pi_a^\bot
   \ket{\varphi^\bot}_{\rm{A}}
   \bra{\varphi^\bot}_{\rm{A}}
 \right)
$.
Combining this relation with the definition of $J_{\{\Pi_a\}}(B|A)$, one obtains
$J_{\{\Pi_a\}}(B|A)=
 J^{\rm{o}}_{\{\Pi_a^\bot\}}(B|A)
$.
Note that
$J(B|A)=
 \sup_{\{\Pi_a\}}
 \left[
   J_{\{\Pi_a\}}(B|A)
 \right]
$,
then it follows that $J^{\rm{o}}(B|A)=J(B|A)$. Thus $J(B|A)$ is preserved.

Consider now the case where the UNOT operation is applied to $B$. Note that the UNOT gate preserves the entropy of any single spin state, in particular when the UNOT operation is applied to $B$, both $S(B)$ and $S(B|a)$ are preserved. Thus, since
$J_{\{\Pi_a\}}(B|A)=
 S(B)-
 \sum_a{
   p_aS(B|a)
 }
$,
the UNOT gate preserves $J_{\{\Pi_a\}}(B|A)$. Therefore $J(B|A)$ is preserved.

\subsection{Proof of result (ii)}

Consider the situation where $\rho$ has no discord, i.e., $\delta(B|A)=0$. Thus $\rho$ is a quantum-classical state, which takes the form
$\rho=
 \sum_{i=0}^1{
   p_i
   \ket{i}_{\rm{A}}
   \bra{i}_{\rm{A}}
   \otimes
   \rho_i^{\rm{B}}
 }
$,
with $\{\ket{i}\}_{i=0,1}$ being an orthogonal basis. The mutual information between $A$ and $B$ is given by
$I(A,B)=
 S\left(
   \sum_{i=0}^1{
     p_i
     \rho_i^{\rm{B}}
   }
 \right)-
 \sum_{i=0}^1{
   p_i
   S(\rho_i^{\rm{B}})
 }
$.

If we apply a UNOT gate to $A$, we transform the state to
$\rho^{\rm{o}}=
 \sum_{i=0}^1{
   p_i
   \ket{i\oplus1}_{\rm{A}}
   \bra{i\oplus1}_{\rm{A}}
   \otimes
   \rho_i^{\rm{B}}
 }
$.
Clearly, the mutual information of $\rho^{\rm{o}}$ is still $I(A,B)$. If we apply a UNOT gate to $B$, both
$S\left(
   \sum_{i=0}^1{
     p_i
     \rho_i^{\rm{B}}
   }
 \right)
$
and $S(\rho_i^{\rm{B}})$ are preserved. Thus $I(A,B)$ remains unchanged.

\section{
  Experimental realization of operations and measurements
  \label{sec:Ops}
}

\subsection{Microwave}

We manipulate the internal states of \Yb by microwave operations which drive transitions
$\ket0_{\rm{A}}
 \leftrightarrow
 \ket{n=1,2,3}_{\rm{A}}
$
resonantly, allowing synthesis of the unitary operations
\begin{equation}
  \label{eq:Rn}
  R_n(\chi,\phi)=
  \exp\left[
    -i\frac\chi2
    \left(
      e^{-i\phi}
      \ket{n}_{\rm{A}}
      \bra0_{\rm{A}}+
      \rm{h.c.}
    \right)
  \right]
\end{equation}
for arbitrary $\chi$ and $\phi$.

Population of the internal state $\ket0_{\rm{A}}$ can be directly measured by standard fluorescence detection and detection error correction~\cite{ChaoShen12}. Population of any other internal state $\ket{n}_{\rm{A}}$ can be measured by first transferring it to $\ket0_{\rm{A}}$ via the microwave operation $R_n(\pi,\phi)$.

\subsection{Red sideband}

We manipulate the external motional degrees of freedom via the red sideband, a standard Raman operation~\cite{Leibfried03} that synthesizes the unitary
\begin{equation}
  \label{eq:R-}
  R_-(\chi,\phi)=
  \exp\left[
    \frac\chi2
    \left(
      e^{-i\phi}
      \sigma_+a-
      e^{i\phi}
      \sigma_-a^\dag
    \right)
  \right].
\end{equation}
Here,
$\sigma_+=
 \ket2_{\rm{A}}
 \bra0_{\rm{A}}
$ and
$\sigma_-=
 \ket0_{\rm{A}}
 \bra2_{\rm{A}}
$,
while $a$ and $a^\dag$ are the annihilation and creation operators with respect to system $B$. The red sideband operation drives transitions
$\ket0_{\rm{A}}
 \ket{m+1}_{\rm{B}}
 \leftrightarrow
 \ket2_{\rm{A}}
 \ket{m}_{\rm{B}}
$
with transition rates depending on $m=0,1,2,\cdots$.

\begin{table}
  \caption{
    \textbf{Pulse sequences of FLIP and SWAP operations.}
    Where $\alpha$ and $\gamma$ in the SWAP operation satisfy
    $\sin(\pi/\sqrt2)
     \cos\alpha=
     \sin(\chi/4)
    $,
    $\tan(\pi/\sqrt2)
     \cos(\phi-\gamma)=
     \tan(\chi/4)
    $.\\
  }
  \begin{indented}
  \item[]
  \begin{tabular}{c|l}
    \hline
    Operation & Sequence \\\hline
    $R_{\rm{FLIP}}(\phi)$ &
    $\begin{array}l
       R_-(\pi/2,\phi)\\
       R_-(\pi/\sqrt2,\phi+\pi/2)\\
       R_-(\pi/2,\phi)
     \end{array}
    $ \\\hline
    $R_{\rm{SWAP}}(\chi,\phi)$ &
    $\begin{array}l
       R_-(\pi/\sqrt2,\gamma)\\
       R_-(\sqrt2\pi,2\alpha+\gamma)\\
       R_-(\pi/\sqrt2,\gamma)
     \end{array}
    $ \\\hline
  \end{tabular}
  \end{indented}
  \label{tab:Red}
\end{table}

\begin{figure}
  \centering
  \includegraphics[width=\columnwidth]{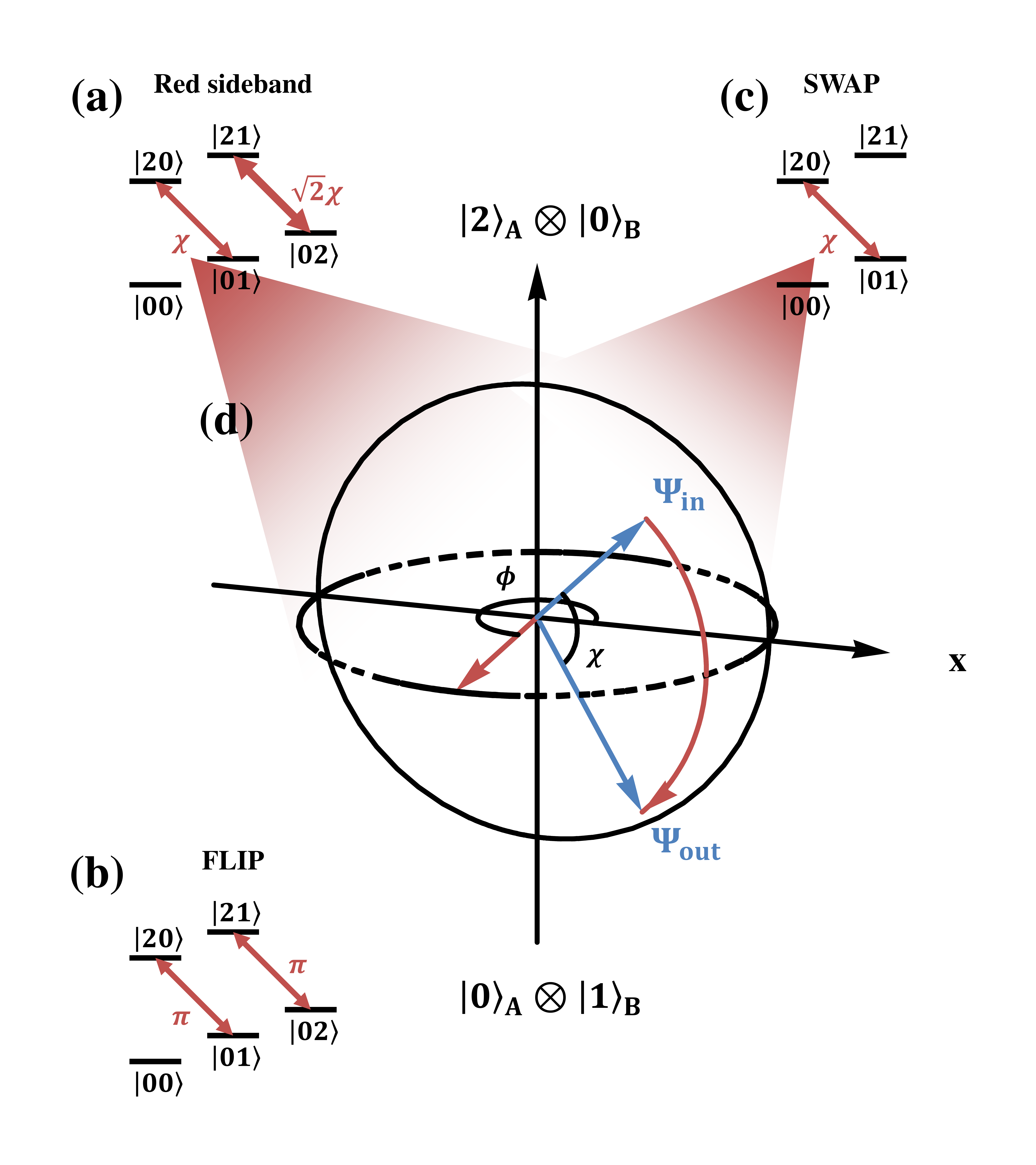}
  \caption{
    \textbf{Effects of external levels involved operations.}
    (a)(b)(c) The effects of the red sideband, FLIP and SWAP operations on low motional energy levels.
    (d) The graphic description of transition
    $\ket0_{\rm{A}}
     \ket1_{\rm{B}}
     \leftrightarrow
     \ket2_{\rm{A}}
     \ket0_{\rm{B}}
    $
    caused by $R_-(\chi,\phi)$ or $R_{\rm{SWAP}}(\chi,\phi)$.
  }
  \label{fig:Red}
\end{figure}

\subsection{FLIP}

The population of
$\ket0_{\rm{A}}
 \ket1_{\rm{B}}
$
can also be probed by measuring the population of $\ket0_{\rm{A}}$, after suitable pre-processing. This is done by first developing a FLIP operation that instigates $\pi$ transitions for both
$\ket0_{\rm{A}}
 \ket1_{\rm{B}}
 \leftrightarrow
 \ket2_{\rm{A}}
 \ket0_{\rm{B}}
$ and
$\ket0_{\rm{A}}
 \ket2_{\rm{B}}
 \leftrightarrow
 \ket2_{\rm{A}}
 \ket1_{\rm{B}}
$~\cite{Gulde03,Monz09,JunhuaZhang16}.
In our protocol, motional levels with $m\geq2$ are unpopulated. Thus measuring the population of $\ket2_{\rm{A}}$ after application of the FLIP operation reveals the the population of
$\ket0_{\rm{A}}
 \ket1_{\rm{B}}
$.

The FLIP operation is composed of 3 red sideband pulses as shown in table~\ref{tab:Red}. Its effect shown in figure~\ref{fig:Red}(b) can be described as following
\begin{eqnarray}
  \label{eq:FLIP}
  \eqalign{
    R_{\rm{FLIP}}(\phi)
    \ket0_{\rm{A}}
    \ket0_{\rm{B}}=
  & \ket0_{\rm{A}}
    \ket0_{\rm{B}}\\
    R_{\rm{FLIP}}(\phi)
    \ket0_{\rm{A}}
    \ket1_{\rm{B}}=
  & \ket2_{\rm{A}}
    \ket0_{\rm{B}}
    e^{
     -i\left(
        \phi+\frac\pi{2\sqrt2}
      \right)
    }\\
    R_{\rm{FLIP}}(\phi)
    \ket0_{\rm{A}}
    \ket2_{\rm{B}}=
  & \ket2_{\rm{A}}
    \ket1_{\rm{B}}
    e^{
     -i\left(
        \phi+\frac\pi2
      \right)
    }\\
    R_{\rm{FLIP}}(\phi)
    \ket1_{\rm{A}}
    \ket{m}_{\rm{B}}=
  & \ket1_{\rm{A}}
    \ket{m}_{\rm{B}}\\
    R_{\rm{FLIP}}(\phi)
    \ket2_{\rm{A}}
    \ket0_{\rm{B}}=
  & \ket0_{\rm{A}}
    \ket1_{\rm{B}}
    e^{
      i\left(
        \phi+\frac\pi{2\sqrt2}
      \right)
    }\\
    R_{\rm{FLIP}}(\phi)
    \ket2_{\rm{A}}
    \ket1_{\rm{B}}=
  & \ket0_{\rm{A}}
    \ket2_{\rm{B}}
    e^{
      i\left(
        \phi+\frac\pi2
      \right)
    }\\
    R_{\rm{FLIP}}(\phi)
    \ket3_{\rm{A}}
    \ket{m}_{\rm{B}}=
  & \ket3_{\rm{A}}
    \ket{m}_{\rm{B}}
  }
\end{eqnarray}
where $m=0,1,2,\cdots$.

\subsection{SWAP}

To measure the population of general
$\ket{n}_{\rm{A}}
 \ket{m}_{\rm{B}}
$,
we develop a generalized version of the SWAP gate of Ref. \cite{Gulde03}, which is able to drive
$\ket0_{\rm{A}}
 \ket1_{\rm{B}}
$
to arbitrary superposition state of
$\ket0_{\rm{A}}
 \ket1_{\rm{B}}
$ and
$\ket2_{\rm{A}}
 \ket0_{\rm{B}}
$,
while preserving the populations of
$\ket0_{\rm{A}}
 \ket0_{\rm{B}}
$ and
$\ket2_{\rm{A}}
 \ket1_{\rm{B}}
$.
This allows us to transfer the population of any general
$\ket{n}_{\rm{A}}
 \ket{m}_{\rm{B}}
$
to state
$\ket0_{\rm{A}}
 \ket1_{\rm{B}}
$
with combined sequence of microwave and SWAP operations.

We develop the SWAP operation with 3 red sideband pulses as shown in table~\ref{tab:Red}. The figure~\ref{fig:Red}(c) shows the effect of SWAP described as following
\begin{eqnarray}
  \label{eq:SWAP}
  \eqalign{
    R_{\rm{SWAP}}(\chi,\phi)
    \ket0_{\rm{A}}
    \ket0_{\rm{B}}=
  & \ket0_{\rm{A}}
    \ket0_{\rm{B}}\\
    R_{\rm{SWAP}}(\chi,\phi)
    \ket0_{\rm{A}}
    \ket1_{\rm{B}}=
  & \ket0_{\rm{A}}
    \ket1_{\rm{B}}
    \cos\frac\chi2+
    \ket2_{\rm{A}}
    \ket0_{\rm{B}}
    \sin\frac\chi2
    e^{-i\phi}\\
    R_{\rm{SWAP}}(\chi,\phi)
    \ket1_{\rm{A}}
    \ket{m}_{\rm{B}}=
  & \ket1_{\rm{A}}
    \ket{m}_{\rm{B}}\\
    R_{\rm{SWAP}}(\chi,\phi)
    \ket2_{\rm{A}}
    \ket0_{\rm{B}}=
  & \ket0_{\rm{A}}
    \ket1_{\rm{B}}
    \sin\frac\chi2
    e^{i\phi}+
    \ket2_{\rm{A}}
    \ket0_{\rm{B}}
    \cos\frac\chi2\\
    R_{\rm{SWAP}}(\chi,\phi)
    \ket3_{\rm{A}}
    \ket{m}_{\rm{B}}=
  & \ket3_{\rm{A}}
    \ket{m}_{\rm{B}}
  }
\end{eqnarray}
where $m=0,1,2,\cdots$.

\section{
  Preparation of spin pair
  \label{sec:Prep}
}

\begin{table}
  \caption{
    \textbf{Preparation of aligned or anti-aligned spin pairs in 6 directions.}
    We obtain the 5th pair
    $\ket0_{\rm{A}}
     \ket0_{\rm{B}}
    $
    by standard sideband cooling process. Other 11 pairs are generated from
    $\ket0_{\rm{A}}
     \ket0_{\rm{B}}
    $
    with corresponding sequences in the right column.\\
  }
  \begin{indented}
  \item[]
  \begin{tabular}{c|c|l}
    \hline
    Spin pair & Encoded form & Sequence \\\hline
    $\ket{\vec{x}}_{\rm{A}}
     \ket{\vec{x}}_{\rm{B}}
    $ &
    $\frac{
       \ket0_{\rm{A}}+
       \ket1_{\rm{A}}
     }{\sqrt2}
     \frac{
       \ket0_{\rm{B}}+
       \ket1_{\rm{B}}
     }{\sqrt2}
    $ &
    $\begin{array}l
       R_2(\pi/2,\pi/2)\\
       R_-(\pi,0)\\
       R_1(\pi/2,-\pi/2)
     \end{array}
    $ \\\hline
    $\ket{-\vec{x}}_{\rm{A}}
     \ket{-\vec{x}}_{\rm{B}}
    $ &
    $\frac{
       \ket0_{\rm{A}}-
       \ket1_{\rm{A}}
     }{\sqrt2}
     \frac{
       \ket0_{\rm{B}}-
       \ket1_{\rm{B}}
     }{\sqrt2}
    $ &
    $\begin{array}l
       R_2(\pi/2,-\pi/2)\\
       R_-(\pi,0)\\
       R_1(\pi/2,\pi/2)
     \end{array}
    $ \\\hline
    $\ket{\vec{y}}_{\rm{A}}
     \ket{\vec{y}}_{\rm{B}}
    $ &
    $\frac{
       \ket0_{\rm{A}}+
       \ket3_{\rm{A}}
     }{\sqrt2}
     \frac{
       \ket0_{\rm{B}}+
       i\ket1_{\rm{B}}
     }{\sqrt2}
    $ &
    $\begin{array}l
       R_2(\pi/2,0)\\
       R_-(\pi,0)\\
       R_3(\pi/2,-\pi/2)
     \end{array}
    $ \\\hline
    $\ket{-\vec{y}}_{\rm{A}}
     \ket{-\vec{y}}_{\rm{B}}
    $ &
    $\frac{
       \ket0_{\rm{A}}-
       \ket3_{\rm{A}}
     }{\sqrt2}
     \frac{
       \ket0_{\rm{B}}-
       i\ket1_{\rm{B}}
     }{\sqrt2}
    $ &
    $\begin{array}l
       R_2(\pi/2,\pi)\\
       R_-(\pi,0)\\
       R_3(\pi/2,\pi/2)
     \end{array}
    $ \\\hline
    $\ket{\vec{z}}_{\rm{A}}
     \ket{\vec{z}}_{\rm{B}}
    $ &
    $\ket0_{\rm{A}}
     \ket0_{\rm{B}}
    $ & \\\hline
    $\ket{-\vec{z}}_{\rm{A}}
     \ket{-\vec{z}}_{\rm{B}}
    $ &
    $\ket1_{\rm{A}}
     \ket1_{\rm{B}}
    $ &
    $\begin{array}l
       R_2(\pi,0)\\
       R_-(\pi,0)\\
       R_1(\pi,0)
     \end{array}
    $ \\\hline
    $\ket{-\vec{x}}_{\rm{A}}
     \ket{\vec{x}}_{\rm{B}}
    $ &
    $\frac{
       \ket0_{\rm{A}}-
       \ket1_{\rm{A}}
     }{\sqrt2}
     \frac{
       \ket0_{\rm{B}}+
       \ket1_{\rm{B}}
     }{\sqrt2}
    $ &
    $\begin{array}l
       R_2(\pi/2,\pi/2)\\
       R_-(\pi,0)\\
       R_1(\pi/2,\pi/2)
     \end{array}
    $ \\\hline
    $\ket{\vec{x}}_{\rm{A}}
     \ket{-\vec{x}}_{\rm{B}}
    $ &
    $\frac{
       \ket0_{\rm{A}}+
       \ket1_{\rm{A}}
     }{\sqrt2}
     \frac{
       \ket0_{\rm{B}}-
       \ket1_{\rm{B}}
     }{\sqrt2}
    $ &
    $\begin{array}l
       R_2(\pi/2,-\pi/2)\\
       R_-(\pi,0)\\
       R_1(\pi/2,-\pi/2)
     \end{array}
    $ \\\hline
    $\ket{-\vec{y}}_{\rm{A}}
     \ket{\vec{y}}_{\rm{B}}
    $ &
    $\frac{
       \ket0_{\rm{A}}-
       \ket3_{\rm{A}}
     }{\sqrt2}
     \frac{
       \ket0_{\rm{B}}+
       i\ket1_{\rm{B}}
     }{\sqrt2}
    $ &
    $\begin{array}l
       R_2(\pi/2,0)\\
       R_-(\pi,0)\\
       R_3(\pi/2,\pi/2)
     \end{array}
    $ \\\hline
    $\ket{\vec{y}}_{\rm{A}}
     \ket{-\vec{y}}_{\rm{B}}
    $ &
    $\frac{
       \ket0_{\rm{A}}+
       \ket3_{\rm{A}}
     }{\sqrt2}
     \frac{
       \ket0_{\rm{B}}-
       i\ket1_{\rm{B}}
     }{\sqrt2}
    $ &
    $\begin{array}l
       R_2(\pi/2,\pi)\\
       R_-(\pi,0)\\
       R_3(\pi/2,-\pi/2)
     \end{array}
    $ \\\hline
    $\ket{-\vec{z}}_{\rm{A}}
     \ket{\vec{z}}_{\rm{B}}
    $ &
    $\ket1_{\rm{A}}
     \ket0_{\rm{B}}
    $ &
    $\begin{array}l
       R_1(\pi,-\pi/2)
     \end{array}
    $ \\\hline
    $\ket{\vec{z}}_{\rm{A}}
     \ket{-\vec{z}}_{\rm{B}}
    $ &
    $\ket0_{\rm{A}}
     \ket1_{\rm{B}}
    $ &
    $\begin{array}l
       R_2(\pi,\pi/2)\\
       R_-(\pi,0)
     \end{array}
    $ \\\hline
  \end{tabular}
  \end{indented}
  \label{tab:Prep}
\end{table}

We prepare $\rhouu$ or $\rhodu$ by deterministically creating pure spin pairs in 6 directions and equally averaging them. The original form, encoded form and the preparation sequence of each spin pair are shown in table~\ref{tab:Prep}.

\section{
  Tomography of spin pair
  \label{sec:Tomo}
}

\begin{table}
  \begin{indented}
  \item[]
  \caption{
    \textbf{Information of the measurements in the tomography.}
    We measure each population
    $P=
     \bra\psi\rho\ket\psi/2=
     \bra{\bar\psi}
     \bar\rho
     \ket{\bar\psi}
    $
    (see \ref{eq:psi}) in the left column, where $\ket\psi$ and $\ket{\bar\psi}$ are listed in the middle and right column respectively.\\
  }
  \begin{tabular}{l|c|c}
    \hline
    Population & state $\ket\psi$ & Measured state \\\hline
    $P_0=\frac{z_0}2
    $ &
    $\ket{\vec{z}}_{\rm{A}}
     \ket{\vec{z}}_{\rm{B}}
    $ &
    $\frac{
       \ket0_{\rm{A}}-
       i\ket2_{\rm{A}}
     }{\sqrt2}
     \ket0_{\rm{B}}
    $ \\\hline
    $P_1=\frac{z_1}2
    $ &
    $\ket{\vec{z}}_{\rm{A}}
     \ket{-\vec{z}}_{\rm{B}}
    $ &
    $\frac{
       \ket0_{\rm{A}}-
       i\ket2_{\rm{A}}
     }{\sqrt2}
     \ket1_{\rm{B}}
    $ \\\hline
    $P_2=\frac{z_2}2
    $ &
    $\ket{-\vec{z}}_{\rm{A}}
     \ket{\vec{z}}_{\rm{B}}
    $ &
    $\frac{
       \ket1_{\rm{A}}-
       i\ket3_{\rm{A}}
     }{\sqrt2}
     \ket0_{\rm{B}}
    $ \\\hline
    $P_3=\frac{z_3}2
    $ &
    $\ket{-\vec{z}}_{\rm{A}}
     \ket{-\vec{z}}_{\rm{B}}
    $ &
    $\frac{
       \ket1_{\rm{A}}-
       i\ket3_{\rm{A}}
     }{\sqrt2}
     \ket1_{\rm{B}}
    $ \\\hline
    $P_4=
     \frac{z_0+z_1}4+
     \frac{x_{0,1}}2
    $ &
    $\ket{\vec{z}}_{\rm{A}}
     \ket{\vec{x}}_{\rm{B}}
    $ &
    $\frac{
       \ket0_{\rm{A}}-
       i\ket2_{\rm{A}}
     }{\sqrt2}
     \frac{
       \ket0_{\rm{B}}+
       \ket1_{\rm{B}}
     }{\sqrt2}
    $ \\\hline
    $P_5=
     \frac{z_0+z_1}4+
     \frac{y_{0,1}}2
    $ &
    $\ket{\vec{z}}_{\rm{A}}
     \ket{\vec{y}}_{\rm{B}}
    $ &
    $\frac{
       \ket0_{\rm{A}}-
       i\ket2_{\rm{A}}
     }{\sqrt2}
     \frac{
       \ket0_{\rm{B}}+
       i\ket1_{\rm{B}}
     }{\sqrt2}
    $ \\\hline
    $P_6=
     \frac{z_0+z_2}4+
     \frac{x_{0,2}}2
    $ &
    $\ket{\vec{x}}_{\rm{A}}
     \ket{\vec{z}}_{\rm{B}}
    $ &
    $\frac{
       \ket0_{\rm{A}}+
       \ket1_{\rm{A}}-
       i\ket2_{\rm{A}}-
       i\ket3_{\rm{A}}
     }2
     \ket0_{\rm{B}}
    $ \\\hline
    $P_7=
     \frac{z_0+z_2}4+
     \frac{y_{0,2}}2
    $ &
    $\ket{\vec{y}}_{\rm{A}}
     \ket{\vec{z}}_{\rm{B}}
    $ &
    $\frac{
       \ket0_{\rm{A}}+
       i\ket1_{\rm{A}}-
       i\ket2_{\rm{A}}+
       \ket3_{\rm{A}}
     }2
     \ket0_{\rm{B}}
    $ \\\hline
    $P_8=
     \frac{z_0+z_3}4+
     \frac{x_{0,3}}2
    $ &
    $\frac{
       \ket\uparrow_{\rm{A}}
       \ket\uparrow_{\rm{B}}+
       \ket\downarrow_{\rm{A}}
       \ket\downarrow_{\rm{B}}
     }{\sqrt2}
    $ &
    $\begin{array}l
       \frac{
         \ket0_{\rm{A}}-
         i\ket2_{\rm{A}}
       }2
       \ket0_{\rm{B}}+\\
       \frac{
         \ket1_{\rm{A}}-
         i\ket3_{\rm{A}}
       }2
       \ket1_{\rm{B}}
     \end{array}
    $ \\\hline
    $P_9=
     \frac{z_0+z_3}4-
     \frac{x_{0,3}}2
    $ &
    $\frac{
       \ket\uparrow_{\rm{A}}
       \ket\uparrow_{\rm{B}}-
       \ket\downarrow_{\rm{A}}
       \ket\downarrow_{\rm{B}}
     }{\sqrt2}
    $ &
    $\begin{array}l
       \frac{
         \ket0_{\rm{A}}-
         i\ket2_{\rm{A}}
       }2
       \ket0_{\rm{B}}-\\
       \frac{
         \ket1_{\rm{A}}-
         i\ket3_{\rm{A}}
       }2
       \ket1_{\rm{B}}
     \end{array}
    $ \\\hline
    $P_{10}=
     \frac{z_0+z_3}4+
     \frac{y_{0,3}}2
    $ &
    $\frac{
       \ket\uparrow_{\rm{A}}
       \ket\uparrow_{\rm{B}}+
       i\ket\downarrow_{\rm{A}}
       \ket\downarrow_{\rm{B}}
     }{\sqrt2}
    $ &
    $\begin{array}l
       \frac{
         \ket0_{\rm{A}}-
         i\ket2_{\rm{A}}
       }2
       \ket0_{\rm{B}}+\\
       \frac{
         i\ket1_{\rm{A}}+
         \ket3_{\rm{A}}
       }2
       \ket1_{\rm{B}}
     \end{array}
    $ \\\hline
    $P_{11}=
     \frac{z_1+z_2}4+
     \frac{x_{1,2}}2
    $ &
    $\frac{
       \ket\uparrow_{\rm{A}}
       \ket\downarrow_{\rm{B}}+
       \ket\downarrow_{\rm{A}}
       \ket\uparrow_{\rm{B}}
     }{\sqrt2}
    $ &
    $\begin{array}l
       \frac{
         \ket0_{\rm{A}}-
         i\ket2_{\rm{A}}
       }2
       \ket1_{\rm{B}}+\\
       \frac{
         \ket1_{\rm{A}}-
         i\ket3_{\rm{A}}
       }2
       \ket0_{\rm{B}}
     \end{array}
    $ \\\hline
    $P_{12}=
     \frac{z_1+z_2}4+
     \frac{y_{1,2}}2
    $ &
    $\frac{
       \ket\uparrow_{\rm{A}}
       \ket\downarrow_{\rm{B}}+
       i\ket\downarrow_{\rm{A}}
       \ket\uparrow_{\rm{B}}
     }{\sqrt2}
    $ &
    $\begin{array}l
      \frac{
         \ket0_{\rm{A}}-
         i\ket2_{\rm{A}}
       }2
       \ket1_{\rm{B}}+\\
       \frac{
         i\ket1_{\rm{A}}+
         \ket3_{\rm{A}}
       }2
       \ket0_{\rm{B}}
     \end{array}
    $ \\\hline
    $P_{13}=
     \frac{z_1+z_3}4+
     \frac{x_{1,3}}2
    $ &
    $\ket{\vec{x}}_{\rm{A}}
     \ket{-\vec{z}}_{\rm{B}}
    $ &
    $\frac{
       \ket0_{\rm{A}}+
       \ket1_{\rm{A}}-
       i\ket2_{\rm{A}}-
       i\ket3_{\rm{A}}
     }2
     \ket1_{\rm{B}}
    $ \\\hline
    $P_{14}=
     \frac{z_1+z_3}4+
     \frac{y_{1,3}}2
    $ &
    $\ket{\vec{y}}_{\rm{A}}
     \ket{-\vec{z}}_{\rm{B}}
    $ &
    $\frac{
       \ket0_{\rm{A}}+
       i\ket1_{\rm{A}}-
       i\ket2_{\rm{A}}+
       \ket3_{\rm{A}}
     }2
     \ket1_{\rm{B}}
    $ \\\hline
    $P_{15}=
     \frac{z_2+z_3}4+
     \frac{x_{2,3}}2
    $ &
    $\ket{-\vec{z}}_{\rm{A}}
     \ket{\vec{x}}_{\rm{B}}
    $ &
    $\frac{
       \ket1_{\rm{A}}-
       i\ket3_{\rm{A}}
     }{\sqrt2}
     \frac{
       \ket0_{\rm{B}}+
       \ket1_{\rm{B}}
     }{\sqrt2}
    $ \\\hline
    $P_{16}=
     \frac{z_2+z_3}4+
     \frac{y_{2,3}}2
    $ &
    $\ket{-\vec{z}}_{\rm{A}}
     \ket{\vec{y}}_{\rm{B}}
    $ &
    $\frac{
       \ket1_{\rm{A}}-
       i\ket3_{\rm{A}}
     }{\sqrt2}
     \frac{
       \ket0_{\rm{B}}+
       i\ket1_{\rm{B}}
     }{\sqrt2}
    $ \\\hline
  \end{tabular}
  \end{indented}
  \label{tab:Tomo}
\end{table}

To reconstruct the spin pair into a density operator
\begin{equation}
  \label{eq:rho}
  \rho=
  \left(
  \begin{array}{cccc}
          z_0        & x_{0,1}-iy_{0,1} & x_{0,2}-iy_{0,2} & x_{0,3}-iy_{0,3}\\
    x_{0,1}+iy_{0,1} &       z_1        & x_{1,2}-iy_{1,2} & x_{1,3}-iy_{1,3}\\
    x_{0,2}+iy_{0,2} & x_{1,2}+iy_{1,2} &       z_2        & x_{2,3}-iy_{2,3}\\
    x_{0,3}+iy_{0,3} & x_{1,3}+iy_{1,3} & x_{2,3}+iy_{2,3} &       z_3
  \end{array}
  \right)
\end{equation}
with basis
$\ket\uparrow_{\rm{A}}
 \ket\uparrow_{\rm{B}}
$,
$\ket\uparrow_{\rm{A}}
 \ket\downarrow_{\rm{B}}
$,
$\ket\downarrow_{\rm{A}}
 \ket\uparrow_{\rm{B}}
$ and
$\ket\downarrow_{\rm{A}}
 \ket\downarrow_{\rm{B}}
$,
we employ quantum state tomography by measuring $\bra\psi\rho\ket\psi$ with several choices of $\ket\psi$.

In experiment, $\rho$ is encoded as
$\bar\rho=
 \mathcal{M}\rho
 \mathcal{M}^\dag
$.
To interpret $\bra\psi\rho\ket\psi$, we derive the decoding procedure
\begin{equation}
  \label{eq:W}
  \mathcal{W}=
  \left(
  \begin{array}{cccccccc}
    1 & & & & i\\
    & 1 & & & & i\\
    & & 1 & & & & i\\
    & & & 1 & & & & i\\
  \end{array}
  \right)
\end{equation}
which is the inverse of the encoding $\mathcal{M}$. Thus
\begin{equation}
  \label{eq:psi}
  \bra\psi\rho\ket\psi=
  \bra\psi
  \mathcal{W}
  \bar\rho
  \mathcal{W}^\dag
  \ket\psi=
  2\bra{\bar\psi}
  \bar\rho
  \ket{\bar\psi}
\end{equation}
where
$\ket{\bar\psi}=
 \mathcal{W}^\dag
 \ket\psi/
 \sqrt2
$.
Each
$\bra{\bar\psi}
 \bar\rho
 \ket{\bar\psi}
$
is obtained by population measurement with 1000 trials. We conduct this population measurement on 6 pure spin pairs from $\rhouu$ or $\rhodu$ with 17 choices of $\ket\psi$ listed in the middle column of table~\ref{tab:Tomo}. In total, we reconstruct $\rhouu$ or $\rhodu$ with $6\times17\times1000=102000$ trials.

We note that
\begin{equation}
  \label{eq:P}
  \eqalign{
    2P_8=
    \bra{\Phi^+}
    \rho
    \ket{\Phi^+}\\
    2P_9=
    \bra{\Phi^-}
    \rho
    \ket{\Phi^-}\\
    2P_{11}=
    \bra{\Psi^+}
    \rho
    \ket{\Psi^+}\\
    2(1-P_8-P_9-P_{11})=
    \bra{\Psi^-}
    \rho
    \ket{\Psi^-}
  }
\end{equation}
where
$\ket{\Phi^+}=
 \left(
   \ket\uparrow_{\rm{A}}
   \ket\uparrow_{\rm{B}}+
   \ket\downarrow_{\rm{A}}
   \ket\downarrow_{\rm{B}}
 \right)/
 \sqrt2
$,
$\ket{\Phi^-}=
 \left(
   \ket\uparrow_{\rm{A}}
   \ket\uparrow_{\rm{B}}-
   \ket\downarrow_{\rm{A}}
   \ket\downarrow_{\rm{B}}
 \right)/
 \sqrt2
$,
$\ket{\Psi^+}=
 \left(
   \ket\uparrow_{\rm{A}}
   \ket\downarrow_{\rm{B}}+
   \ket\downarrow_{\rm{A}}
   \ket\uparrow_{\rm{B}}
 \right)/
 \sqrt2
$ and
$\ket{\Psi^-}=
 \left(
   \ket\uparrow_{\rm{A}}
   \ket\downarrow_{\rm{B}}-
   \ket\downarrow_{\rm{A}}
   \ket\uparrow_{\rm{B}}
 \right)/
 \sqrt2
$
are Bell states. Thus $P_8$, $P_9$ and $P_{11}$ provide Bell-basis measurements for $\rho$.

\begin{figure*}
  \centering
  \includegraphics[width=\columnwidth]{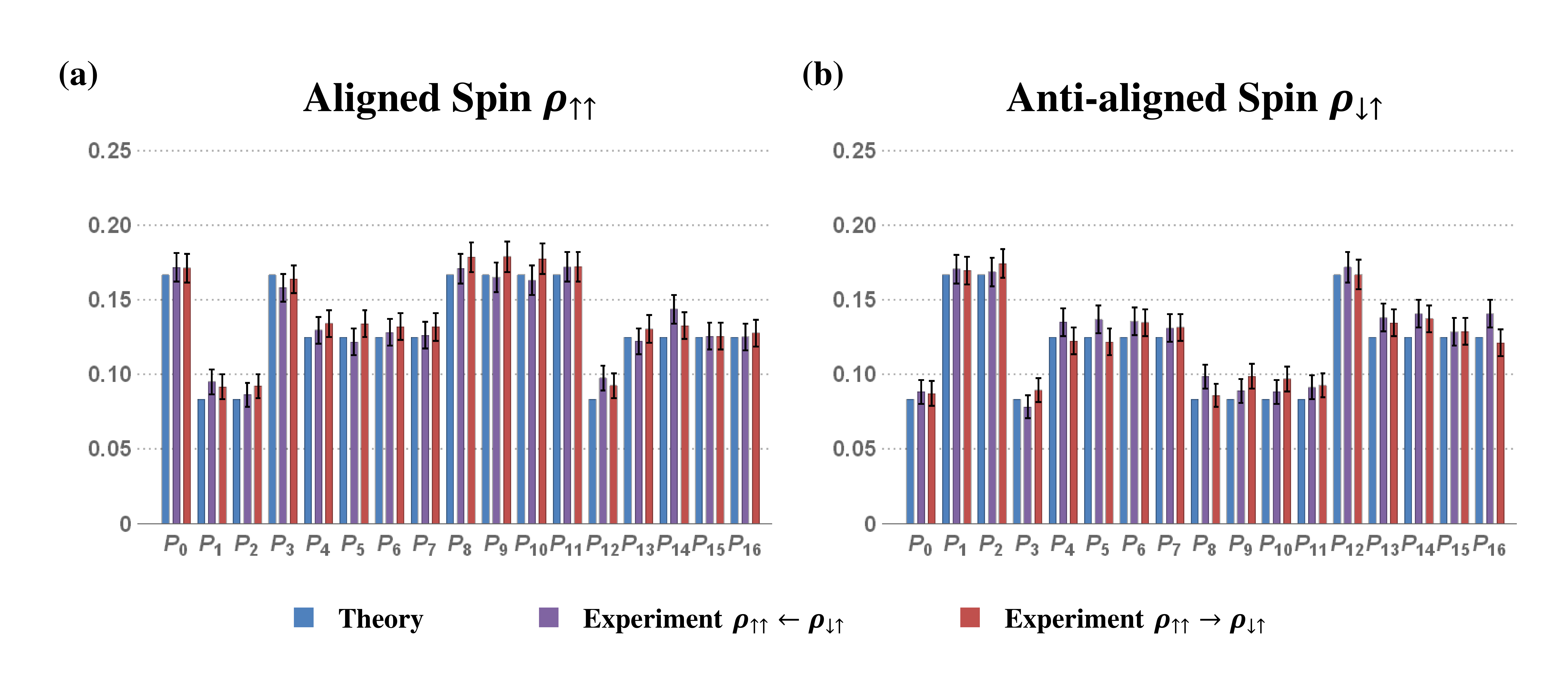}
  \caption{
    \textbf{The comparison between theoretical and experimental results of 17 population measurements} for
    (a) the aligned spin $\rhouu$ and
    (b) the anti-aligned spin $\rhodu$.
    The error bar of each quantity is calculated with confidence level of 95\%.
  }
  \label{fig:Tomo}
\end{figure*}

\subsection{Density operator reconstruction}

We reconstruct the most probable density operator of $\rhouu$ or $\rhodu$ from the experimental data of $P_j$ exhibited in figure~\ref{fig:Tomo}, subject to constraint that it is normalized ($z_0+z_1+z_2+z_3=1$) and non-negative definite. We determine the maximum likelihood of $\rhouu$ or $\rhodu$ by minimizing
\begin{equation}
  \label{eq:Object}
  \sum_{j=0}^{16}{
    \frac{
      (P_j-P_j^E)^2
    }{2\Delta_j^2}
  }
\end{equation}
where $P_j^E$ is the experimental result of $P_j$, and $\Delta_j$ is the uncertainty of $P_j^E$. This objective function is obtained by maximizing the probability density
\begin{equation}
  \label{eq:f}
  f(\rho)=
  \prod_{j=0}^{16}{
    \frac1{\sqrt{2\pi}\sigma_j}
    \exp\left[
      -\frac{
        (P_j-P_j^E)^2
      }{2\sigma_j^2}
    \right]
  },
\end{equation}
supposing that each variable $P_j$ follows a Gaussian distribution with mean $P_j^E$ and standard deviation $\sigma_j$. Here we set $\Delta_j=1.96\sigma_j$, corresponding to the confidence level of 95\%.

\clearpage

\section*{References}


\begin{thebibliography}{10}
\expandafter\ifx\csname url\endcsname\relax
  \def\url#1{{\tt #1}}\fi
\expandafter\ifx\csname urlprefix\endcsname\relax\def\urlprefix{URL }\fi
\providecommand{\eprint}[2][]{\url{#2}}

\bibitem{Gisin99}
Gisin N and Popescu S 1999 {\em Phys. Rev. Lett.\/} {\bf 83}(2) 432--435
  \urlprefix\url{http://link.aps.org/doi/10.1103/PhysRevLett.83.432}

\bibitem{Bechmann99}
Bechmann-Pasquinucci H and Gisin N 1999 {\em Phys. Rev. A\/} {\bf 59}(6)
  4238--4248 \urlprefix\url{http://link.aps.org/doi/10.1103/PhysRevA.59.4238}

\bibitem{Werner99}
Bu\ifmmode~\check{z}\else \v{z}\fi{}ek V, Hillery M and Werner R~F 1999 {\em
  Phys. Rev. A\/} {\bf 60}(4) R2626--R2629
  \urlprefix\url{http://link.aps.org/doi/10.1103/PhysRevA.60.R2626}

\bibitem{Martini02}
De~Martini F, Bu{\v{z}}ek V, Sciarrino F and Sias C 2002 {\em Nature\/} {\bf
  419} 815--818

\bibitem{Ricci04}
Ricci M, Sciarrino F, Sias C and De~Martini F 2004 {\em Phys. Rev. Lett.\/}
  {\bf 92}(4) 047901
  \urlprefix\url{http://link.aps.org/doi/10.1103/PhysRevLett.92.047901}

\bibitem{Scarani05}
Scarani V, Iblisdir S, Gisin N and Ac\'{\i}n A 2005 {\em Rev. Mod. Phys.\/}
  {\bf 77}(4) 1225--1256
  \urlprefix\url{http://link.aps.org/doi/10.1103/RevModPhys.77.1225}

\bibitem{Jeffrey06}
Jeffrey E~R, Altepeter J~B, Colci M and Kwiat P~G 2006 {\em Phys. Rev. Lett.\/}
  {\bf 96}(15) 150503
  \urlprefix\url{http://link.aps.org/doi/10.1103/PhysRevLett.96.150503}

\bibitem{Ollivier01}
Ollivier H and Zurek W~H 2001 {\em Phys. Rev. Lett.\/} {\bf 88}(1) 017901
  \urlprefix\url{http://link.aps.org/doi/10.1103/PhysRevLett.88.017901}

\bibitem{Vedral01}
Henderson L and Vedral V 2001 {\em Journal of Physics A: Mathematical and
  General\/} {\bf 34} 6899
  \urlprefix\url{http://stacks.iop.org/0305-4470/34/i=35/a=315}

\bibitem{MileGu12}
Gu M, Chrzanowski H~M, Assad S~M, Symul T, Modi K, Ralph T~C, Vedral V and Lam
  P~K 2012 {\em Nature Physics\/} {\bf 8} 671--675

\bibitem{Leibfried03}
Leibfried D, Blatt R, Monroe C and Wineland D 2003 {\em Rev. Mod. Phys.\/} {\bf
  75}(1) 281--324
  \urlprefix\url{http://link.aps.org/doi/10.1103/RevModPhys.75.281}

\bibitem{Casanova11}
Casanova J, Sab\'{\i}n C, Le\'on J, Egusquiza I~L, Gerritsma R, Roos C~F,
  Garc\'{\i}a-Ripoll J~J and Solano E 2011 {\em Phys. Rev. X\/} {\bf 1}(2)
  021018 \urlprefix\url{http://link.aps.org/doi/10.1103/PhysRevX.1.021018}

\bibitem{XiangZhang15}
Zhang X, Shen Y, Zhang J, Casanova J, Lamata L, Solano E, Yung M~H, Zhang J~N
  and Kim K 2015 {\em Nature communications\/} {\bf 6}

\bibitem{Cover06}
Cover T and Thomas J 2006 {\em Elements of Information Theory\/} A
  Wiley-Interscience publication (Wiley) ISBN 9780471748816
  \urlprefix\url{https://books.google.co.jp/books?id=EuhBluW31hsC}

\bibitem{Nielsen10}
Nielsen M and Chuang I 2010 {\em Quantum Computation and Quantum Information\/}
  (Cambridge University Press) ISBN 9780511992773
  \urlprefix\url{https://books.google.com.hk/books?id=JRz3jgEACAAJ}

\bibitem{YangchaoShen17}
Shen Y, Zhang X, Zhang S, Zhang J~N, Yung M~H and Kim K 2017 {\em Phys. Rev.
  A\/} {\bf 95}(2) 020501
  \urlprefix\url{http://link.aps.org/doi/10.1103/PhysRevA.95.020501}

\bibitem{Gulde03}
Gulde S, Riebe M, Lancaster G~P, Becher C, Eschner J, H{\"a}ffner H,
  Schmidt-Kaler F, Chuang I~L and Blatt R 2003 {\em Nature\/} {\bf 421} 48--50

\bibitem{Monz09}
Monz T, Kim K, H\"ansel W, Riebe M, Villar A~S, Schindler P, Chwalla M,
  Hennrich M and Blatt R 2009 {\em Phys. Rev. Lett.\/} {\bf 102}(4) 040501
  \urlprefix\url{http://link.aps.org/doi/10.1103/PhysRevLett.102.040501}

\bibitem{JunhuaZhang16}
{Zhang} J, {Um} M, {Lv} D, {Zhang} J~N, {Duan} L~M and {Kim} K 2016 {\em ArXiv
  e-prints\/} (\textit{Preprint} \eprint{1611.08700})

\bibitem{Holevo73}
Holevo A~S 1973 {\em Problemy Peredachi Informatsii\/} {\bf 9} 3--11

\bibitem{Modi14}
Modi K 2014 {\em Open Systems \& Information Dynamics\/} {\bf 21} 1440006
  \urlprefix\url{http://www.worldscientific.com/doi/abs/10.1142/S123016121440006X}

\bibitem{Weedbrook13}
{Weedbrook} C, {Pirandola} S, {Thompson} J, {Vedral} V and {Gu} M 2013 {\em
  ArXiv e-prints\/} (\textit{Preprint} \eprint{1312.3332})

\bibitem{Niset07}
Niset J, Ac\'{\i}n A, Andersen U~L, Cerf N~J, Garc\'{\i}a-Patr\'on R,
  Navascu\'es M and Sabuncu M 2007 {\em Phys. Rev. Lett.\/} {\bf 98}(26) 260404
  \urlprefix\url{https://link.aps.org/doi/10.1103/PhysRevLett.98.260404}

\bibitem{ChaoShen12}
Shen C and Duan L~M 2012 {\em New Journal of Physics\/} {\bf 14} 053053
  \urlprefix\url{http://stacks.iop.org/1367-2630/14/i=5/a=053053}

\end{thebibliography}

\end{document}